# EXCEPTIONS IN BUSINESS PROCESSES IN RELATION TO OPERATIONAL PERFORMANCE

*Remco Dijkman[*], Geoffrey van IJzendoorn[2], Oktay Turetken[1], Meint de Vries[3]*

[1] *Eindhoven University of Technology, Eindhoven, The Netherlands*

r.m.dijkman@tue.nl, o.turetken@tue.nl

[2] *Deloitte, Utrecht, The Netherlands*

GvanIJzendoorn@deloitte.nl

[3] *Accenture, Amsterdam, The Netherlands*

meint.de.vries@accenture.com

**Abstract**

*Business process models describe the way of working in an organization. Typically, business process models distinguish between the normal flow of work and exceptions to that normal flow. However, they often present an idealized view. This means that unexpected exceptions – exceptions that are not modelled in the business process model – can also occur in practice. This has an effect on the efficiency of the organization, because information systems are not developed to handle unexpected exceptions. This paper studies the relation between the occurrence of exceptions and operational performance. It does this by analyzing the execution logs of business processes from five organizations, classifying execution paths as normal or exceptional. Subsequently, it analyzes the differences between normal and exceptional paths. The results show that exceptions are related to worse operational performance in terms of a longer throughput time and that unexpected exceptions relate to a stronger increase in throughput time than expected exceptions.*

# 1    Introduction

Organizations often describe their flow of operations in terms of business process models. Tools for operations support, such as information systems, forms, and manuals, are then developed based on these business process models. However, in practice exceptions can occur that disrupt the normal flow of operations (Reichert and Weber, 2012b). The tools for operations support, which have been defined for the normal flow of operations, cannot be expected to provide optimal support in case of an exception. For that reason, we expect that the performance of business processes deteriorates when exceptions occur.

For example, consider a bank that has a process for applying for a bank account. The application process is supported by several forms that a new customer has to fill out, manuals that explain to the bank employee what to do during the various steps of the application procedure, and information systems for recording customer details and activating the bank account. There is likely to be a normal flow through the process that most customers follow and that can be dealt with efficiently, because the employees have been trained for and have experience with that normal flow, and because all forms, work instructions, and information systems are optimized for it. It is also likely that the bank pre-defined certain exceptions to the normal flow that are expected to occur incidentally, such as the situation in which a customer fills out a form incorrectly. Such exceptions may lead to delays in the process and a decrease in quality and customer satisfaction, simply because more work is required, but also because employees have less experience with these situations, depending on how frequently they occur. We will show in this paper that exceptions that were *not* considered when designing a process, form, or information system are also surprisingly frequent. Such exceptions can be expected to further decrease the efficiency and quality of the work. Just remember the time that a helpdesk employee told you: "I am sorry Sir/Madam, but our system does not allow for that."

The goal of this paper is to investigate the relation between different types of exceptions in a business process and the performance of that business process. To achieve this goal, the paper provides a classification of the different types of exceptions that can occur during the execution of a business process. Subsequently, it presents a statistical analysis of the relations between the different types of exceptions and the performance of a business process – in particular the throughput time. To the best of our knowledge, such an analysis has not been made before.

This has important practical implications. First, there exist many papers that propose solutions for handling exceptions (Schonenberg, Mans et al., 2008), based on the premise that exceptions are a fact of life and deteriorate process performance when not properly supported. This study validates that premise. Second, this study provides a theoretical basis for showing organizations how different types of exceptions should be supported in practice.

The study is conducted by investigating five processes in five different companies and a total of nearly 70,000 customer cases that have been executed for these processes. Of these customer cases more than 40,000 contained some form of exception. These numbers are sufficient to do a statistical analysis for each of the business processes. Subsequently we aggregate the results of the statistical analyses in a qualitative manner.

Against this background, the remainder of this paper is structured as follows. Section 2 presents the background of the research by providing a conceptual background on exceptions and an overview of related work. Section 3 discusses the research methodology. Section 4 and 5 present the analysis of the data and the results of the study, and Section 6 presents the conclusions.

# 2    Background

A business process is a structured, measured set of activities designed to produce a specific output for a particular customer (Davenport, 1993). Business processes are the structure by which an organization does what is necessary to produce value for its customers. A business process model is a (graph-ical) representation of a business process. Figure 1 shows an example of a business process model, in the notation of automated process discovery tool 'Disco'[1]. The process starts at the 'play' icon and stops at the 'stop' icon. Rounded rectangles represent tasks that can be executed. An arrow represents that, if the task at the source of the arrow is completed, the task at the target of the arrow can be started. When multiple arrows leave a task, this represents a choice between the target tasks. For example, after 'register appeal' there is a choice between 'request revision', 'register documents' and 'reject appeal'. Multiple incoming arcs to a task also indicate an (exclusive) choice that triggers the task. For instance, 'request revision' task can be triggered by 'request appeal' or 'register documents'.

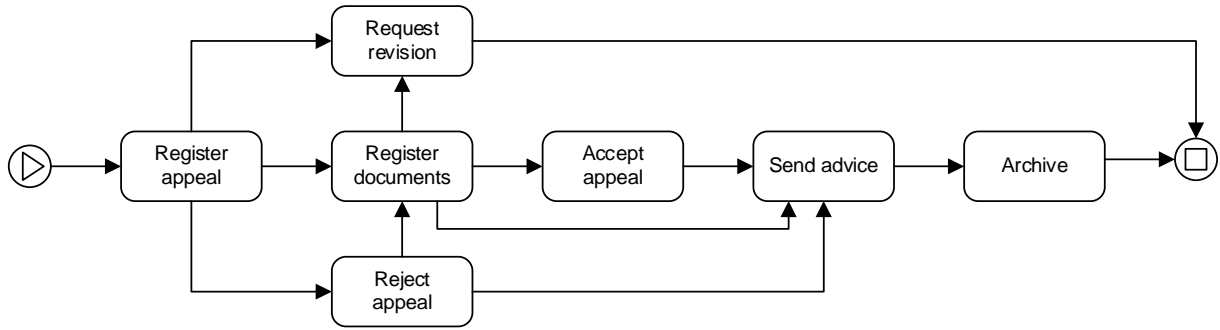

**Figure 1. A Business Process Model.**

Each customer case follows a certain path through the process. For example, for the process above, a possible path is: 'register appeal', 'reject appeal', 'send advice', and 'archive'. Another possible path is 'register appeal', 'request revision'.

Different paths through the process may lead to a different outcome of the process. For example, a path that contains 'request revisions' has a different outcome than a path that contains 'accept appeal'. We call a collection of paths that leads to the same outcome a *scenario*. The process from Figure 1 has three different scenarios: a scenario for accepting an appeal, a scenario for rejecting an appeal and a scenario for requesting revisions. However, there are different paths for each scenario. For example, an appeal can be rejected via the path 'register appeal', 'reject appeal', 'send advice', 'archive', but also via the path 'register appeal', 'reject appeal', 'register documents', 'send advice', 'archive'.

Different paths also have a different frequency in which they are being taken by customer cases. Each scenario has a most frequently followed path, which will be called the *normal flow* of the scenario in the remainder of this paper. An exception is a disruption of the normal flow of operations (Reichert and Weber, 2012b). While there exist business process modelling languages that have notational elements for modelling exceptions, such as the Business Process Model and Notation (Object Management Group, 2011), such notational elements are not frequently used (zur Muehlen and Recker, 2008). Consequently, it is often hard to determine which paths through a process represent normal behavior and which paths represent exceptional behavior (Grigori, Casati, Dayal and Shan, 2001). In this paper we define an *exception* as a path through a process that is not the normal flow of a *scenario*. This position is based on the observation that non-normal flows are relatively rare, as we will show in Section 4, such that they are exceptional in the dictionary sense of the word. We should note that, this definition adopts a rather restricted view of process exceptions in the sense that it focuses only on the *control-flow* aspect of processes and does not cover exceptions regarding other aspects such as resources (roles, actors, etc.) or information.

A clearly identifiable form of exception is an exception that causes a customer case to follow a path that is not represented by the process model. For example, if a customer follows the path 'register ap-peal', 'request revision', 'accept appeal', 'send advice', 'archive' through the business process from Figure 1, this would be an exception to the normal flow of operations. Consequently, we define an *unexpected exception* as a path through a process that is not described by the process model and an *expected exception* as an exception that is described by the process model.

---
[1] http://fluxicon.com/disco/

Looking in more detail at the paths that are taken in the process outlined above, a more fine-grained distinction between exceptions can be made. Considering a normal flow A-B-C-D, we define an exception as:

- an early exit, when we remove an activity (in our case D) from the end of the path;
- a late exit, when we add an activity to the end of the path;
- an early entry, when we remove an activity (in our case A) from the start of the path;
- a late entry, when we add an activity to the start of the path;
- a repeat, when a single activity is repeated (for example: A-B-C-C-D);
- a step back, when a sequence of multiple activities is repeated (for example: A-B-C-B-C-D);
- an add, when an activity is added to the normal flow, but it is not a late exit, late entry, repeat, or step back; and
- a skip, when an activity is removed from the normal flow, but it is not an early entry, or an early exit.

There is a large body of literature on exception handling in business processes. Various papers discuss the topic from different perspectives. First, there are papers that conceptualize the notion of exception and related notions (Curbera, Khalaf, Leymann and Weerawarana, 2003; Eder and Liebhart, 1998; Strong, 1997) and papers that present the various types of exceptions that can occur during the execution of a business process (Russell, van der Aalst and ter Hofstede, 2006; Weber, Reichert and Rinderle-Ma, 2008). From a technical perspective, there exists a line of research on developing techniques for detecting whether an exception has occurred and even predicting whether an exception will occur for a certain customer case (Grigori, Casati, Castellanos, Dayal, Sayal and Shan, 2004; Grigori, Casati, Dayal and Shan, 2001; Hwang, Ho and Tang, 1999; Rozinat and van der Aalst, 2008). In addition to that there exists a line of research on developing techniques for handling exceptions when they occur (Adams, ter Hofstede, Edmond and van der Aalst, 2006; Zeng, Lei, Jeng, Chung and Benatallah, 2005; Hagen and Alonso, 2000; Rinderle and Reichert, 2006; Gottschalk, Wagemakers et al., 2009; van der Aalst, 2009; Hallerbach, Bauer et al. 2010; Meerkamm, 2012; Reichert and Dadam, 1998; Reichert and Weber, 2012a). Schonenberg, Mans et al. (2008) provide a literature survey of those tools and techniques. To the best of our knowledge, there is no existing work on determining the relations between exceptions and the performance of a business process. This paper aims to fill that gap.

## 3 Research Methodology

Based on the discussion from the previous section, the overall research goal can be refined into hypotheses. In particular, the following hypotheses follow from the discussion.

H1. Paths through a process that involve an exception have a longer throughput time than paths through a process that do not involve an exception.

H2. Paths that involve an unexpected exception have a longer throughput time than paths that involve an expected exception.

H3. Paths that involve an exception that adds activities to the normal flow have a longer throughput time than the normal flow.

H4. Paths that involve an exception that removes activities from the normal flow have a shorter throughput time than the normal flow.

Hypothesis 1 and 4 seem to contradict each other. However, note that hypothesis 1 concerns all exceptions, while hypothesis 4 only concerns a subset of the exceptions; those exceptions that remove activities from the normal flow. If this subset of exceptions is relatively small, it is still possible that, overall, exceptions are related to an increase in throughput time.

For reasons of availability of data, this paper focuses on the operational performance in terms of throughput time. In future work we aim to study other key performance indicators, such as customer satisfaction.

The following research steps were followed to investigate the hypotheses. First, we selected the processes to be analyzed and gathered process related data including the process models, where available, and process instance data (cases). We selected five processes from five companies operating in different business domains. For each of the five processes, we gathered information on the customer cases that flowed through the process during a certain time period.

Second, using process-mining techniques, we analyzed the cases for each process to determine the paths that are followed through the process and corresponding throughput time. This information allows us to determine whether the customer case contained any exceptions and to compare the throughput time between customer cases. For two of the five processes, we also have the original process model, which allows us to investigate differences between expected and unexpected exceptions.

Finally, we performed a set of statistical analyses with the data originated from the previous step to test our hypotheses. First, we performed the analysis on the customer cases of each process separately, in which we determined significant differences between the throughput times of customer cases, depending on the types of exception that they contain. Based on this analysis, we established per process and per type of exception whether it is associated with a significantly longer or shorter throughput time, both compared to the normal flow and compared to the flow of work that is described by the process model. Second, we aggregated the results over the processes, using qualitative arguments. When the same type of exception has the same type of relation for multiple processes, it is considered that that type of exception indeed follows that type of relation in the general case.

## 4   Descriptive statistics

Table 1 shows the descriptive statistics of the processes that are investigated, including the type of process used, the start date of the measurements, the end date of the measurements, the number of customer cases measured and the average throughput times of the customer cases and the standard deviation of the throughput times.

| process | start date | end date | nr. cases | avg. throughput | std. throughput |
|---|---|---|---|---|---|
| P1 - appeals | 24-10-2006 | 20-10-2008 | 1,268 | 41 weeks | 31 weeks |
| P2 - change control | 09-03-2006 | 22-10-2007 | 4,014 | 52 days | 55 days |
| P3 - incident management | 08-09-2008 | 19-01-2010 | 41,902 | 19 days | 21 days |
| P4 - loan application | 01-10-2011 | 14-03-2012 | 13,085 | 8 days | 12 days |
| P5 - incident management | 05-05-2011 | 23-05-2012 | 7,554 | 12 days | 29 days |

**Table 1. Descriptive statistics of processes.**

The first process is an appeals process at the municipality of Eindhoven. This process concerns appeals by citizens for a revision of a decision that was made by the municipality at an earlier stage. There are five possible outcomes of this process. The appeal can be accepted, rejected, withdrawn by the citizen, sent on to another department, or deferred awaiting a revision of the request from the citizen. These outcomes would seem suitable as scenarios. However, the data shows that it is possible to have 'multiple outcomes' at the same time, due to the fact that the different parties involved can take decisions independently. For example, it is possible that the citizen withdraws the appeal *and* the municipality rejects the appeal. Since it is impossible to decide on the exact outcome in such cases, we consider these outcomes separately and count them as different scenarios.

The second process is a change control process at a consumer electronics manufacturing company. This process concerns evaluating whether change requests should be incorporated into a product. From a process perspective, there is only one possible scenario in this process: the scenario in which the change request is handled. The process does not have different paths for change requests that are rejected and change requests that are accepted. Therefore, we cannot distinguish these as different scenarios. For this process, we originally had nearly two years of data. However, halfway the measurement period, there was a change to the process, such that customer cases that finished before that date were handled differently from customer cases that were finished on or after that date. To not pollute the measurements with this, we only included customer cases from the most recent period.

| process | scenario | type of path | nr. of paths | avg. nr. of cases per path |
|---|---|---|---|---|
| P1 | S1-Accepted | normal | 1 | 498 |
| | | exceptions | 7 | 40 |
| | S2-Withdrawn | normal | 1 | 78 |
| | | exceptions | 9 | 11 |
| | S3-Rejected | normal | 1 | 57 |
| | | exceptions | 8 | 13 |
| | S4-Deferred | normal | 1 | 47 |
| | | exceptions | 9 | 5 |
| | S5-Sent on | normal | 1 | 27 |
| | S6-Withdrawn & Rejected | normal | 1 | 13 |
| | | exceptions | 6 | 4 |
| | S7-Deferred & Rejected | normal | 1 | 4 |
| | | exceptions | 1 | 1 |
| | S8-Deferred & Withdrawn | normal | 1 | 1 |
| P2 | S1-Handled | normal | 1 | 2400 |
| | | exceptions | 44 | 37 |
| P3 | S1-Handled | normal | 1 | 9913 |
| | | exceptions | 934 | 34 |
| | S2-Cancelled | normal | 1 | 333 |
| | | exceptions | 52 | 7 |
| P4 | S1-Rejected | normal | 1 | 5719 |
| | | exceptions | 33 | 70 |
| | S2-Accepted | normal | 1 | 1535 |
| | | exceptions | 30 | 24 |
| | S3-Cancelled | normal | 1 | 1132 |
| | | exceptions | 20 | 84 |
| P5 | S1-Handled by helpdesk | normal | 1 | 1749 |
| | | exceptions | 29 | 5 |
| | S2-Handled by others | normal | 1 | 524 |
| | | exceptions | 2247 | 2 |

**Table 2. Scenarios per process**

The third process is an incident management process at an American company. This process concerns handling incidents that are reported to the company, from the moment that they are submitted until the moment that they are handled. This process has two possible outcomes and corresponding scenarios. Either the incident is handled, or the incident report is cancelled.

The fourth process is a loan application process at a Dutch financial institution (van Dongen, 2012). This process concerns applications for loans, from the moment that they are submitted by the client until the moment that they are accepted, rejected, or cancelled by the customer. This process has three possible outcomes and corresponding scenarios. The loan can be accepted, rejected, or the loan application can be cancelled.

The fifth process is an incident management process at Volvo (Steeman, 2013). Similar to the incident management process from the third process, this process also concerns handling incidents, from the moment that they are submitted until the moment that they are handled. However, the number and flow of tasks that are used in this process varies from those in the third process. This process has two possible outcomes and corresponding scenarios. Either the incident report is handled till completion by the helpdesk, or it is handled by the second or third line.

Table 2 shows the different scenarios that occur in each of the processes that we investigate. For each scenario, it also shows the number of possible paths through that scenario, distinguishing between the normal flow of the scenario (of which there is one by definition) and exceptions to that normal flow (of which there are typically more than one). It then shows the average number of customer cases through each path.

The information in Table 2 shows that an exceptional path on average is taken far less frequently than a normal flow. This is more clearly illustrated in the graph in Figure 2, which shows the 15 most frequently taken paths through each process and the percentage of customer cases that take these paths. The figure shows that each process has one path that is by far the most traversed path. Except for the second process, the next most traversed path takes less than half of the number of customer cases that the first path takes. After that there are a number of paths that are followed by a distinguishable number of customer cases. The number of customer cases per path then rapidly decreases and the far majority of the paths (98.5%) take less than 1% of the customer cases.

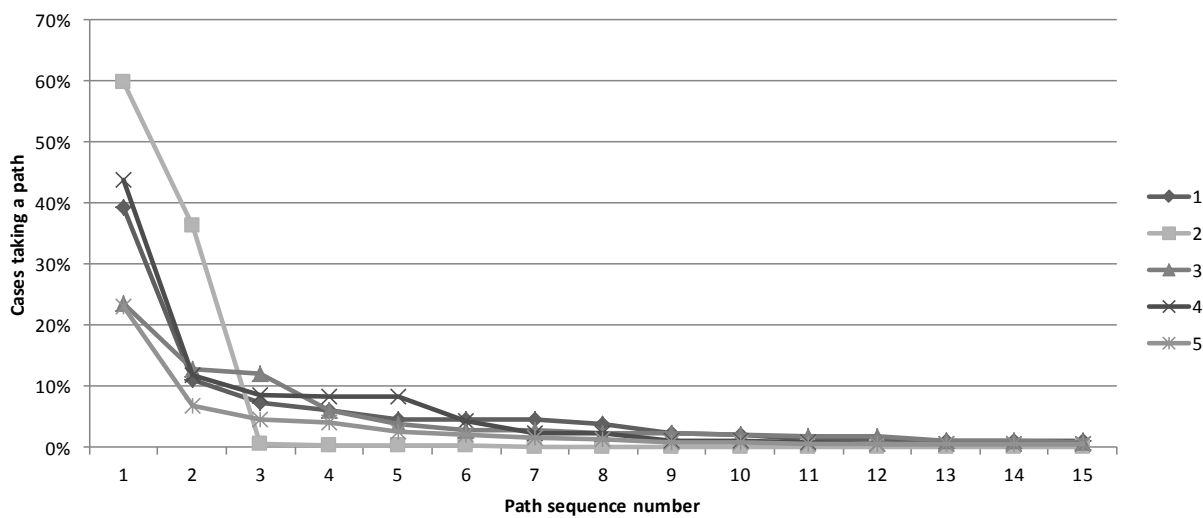

**Figure 2. Frequency with which the 15 most popular paths are taken**

Figure 2 also shows that there are exceptional paths that are followed as frequently as the normal flows. Therefore, the precise definition of exceptional path can be debated. Clearly paths that are followed twice as often as other paths are normal flows, while paths that are taken by less than 1% of the customer cases are exceptions. However, the question remains where between these two extremes we consider a path an exceptional path. There exist rules of thumb, such as the elbow principle (Tibshirani, Walther and Hastie, 2001) and the pareto, or 80/20, principle. However, further analysis shows that these rules do not apply here. Using the elbow principle would mean excluding paths that clearly represent common process behavior, such as rejecting loans. Using the pareto principle would mean including a large number of paths that are taken by less than 1% of the customers. Another logical assumption would be that there is a relation between the paths that are described in the process model and exceptions, but that is also not the case. There are paths in the process models that are taken by less than 1% of the customer cases. Moreover, the normal path through process 2 is not in the process model. For these reasons, we used the concept of scenarios to identify normal flows, as explained in Section 2, defining a normal flow as the most frequently taken path to achieve an identifiable result.

For process 2 and process 5 we also have the process models, such that it is possible to determine the difference between expected exceptions – exception paths that are described by the process model – and unexpected exceptions – exception paths that are not described by the process model. Table 3 shows the differences between these two types of exceptions. The table illustrates that there are more unexpected paths than there are expected paths and that each unexpected path is taken by relatively few customer cases. That means that each expected exception occurs less frequently than the normal flow and that each unexpected exception occurs even less frequently.

| process | scenario | type of path | nr. of paths | avg. nr. of cases per path |
|---|---|---|---|---|
| P2 | S1-Handled | normal | 1 | 2400 |
| | | expected | 15 | 102 |
| | | unexpected | 29 | 3 |
| P5 | S1-Handled by helpdesk | normal | 1 | 1749 |
| | | unexpected | 29 | 5 |
| | S2-Handled by others | normal | 1 | 524 |
| | | expected | 118 | 5 |
| | | unexpected | 2129 | 2 |

**Table 3. Expected and unexpected exceptions per process**

Looking in more detail at the different types of exceptions, as they are defined in Section 2, Table 4 shows the relative frequency with which the different types of exceptions occur in the paths through the process.

| process | early entry | late entry | early exit | late exit | repeat | step back | add | skip |
|---|---|---|---|---|---|---|---|---|
| P1 | 0% | 0% | 25% | 8% | 8% | 4% | 35% | 33% |
| P2 | 0% | 0% | 0% | 0% | 20% | 48% | 65% | 26% |
| P3 | 1% | 0% | 8% | 22% | 73% | 48% | 49% | 29% |
| P4 | 0% | 0% | 0% | 19% | 0% | 67% | 30% | 26% |
| P5 | 2% | 96% | 3% | 1% | 11% | 63% | 48% | 1% |

**Table 4. Relative frequency of types of exceptions**

## 5     Analysis and results

In order to investigate the relation of different exception types and the throughput time, we conducted a series of statistical analyses over the customer cases of each process and corresponding scenarios. Such relationships can be analyzed using parametric statistical tests, such as the linear regression, where the exception types can be considered as predictors (independent variables) and the throughput time as the dependent variable. These tests, however, give valid results only if a certain set of assumptions holds (Witte and Witte, 2014), especially the normality assumption (Field, 2013).

The results of our initial analysis showed that there are clear deviations from normality in throughput time for all processes and scenarios (with a Shapiro-Wilk test of normality with $p < 0.01$). This remained as a limitation even after transforming (e.g., log-normalizing) the dependent variable. Therefore, instead of applying parametric statistical tests, we forewent the predictive power of such tests and applied their non-parametric counterparts (Field, 2013). In particular, we used the Kruskal–Wallis H test to evaluate the hypothesis that multiple independent groups come from different populations. In other words, we assessed per exception type if the throughput time for the customer cases with specific types of exceptions is different than the throughput time for the customer cases that belong to the normal flow of a scenario.

We included in our analyses only those customer cases that have at most two exceptions types occurring at the same time and have a group size larger than 25. In doing so, we aimed to preserve accuracy in the inferences from the Kruskal–Wallis H tests and balance the sizes of the groups that are compared.

Table 5 shows the results of the statistical analysis of the relation of exceptions in a process on the throughput time of that process. It also shows the descriptive statistics of the scenarios that were investigated: the number of process instances in the scenario, as well as the average throughput time of the scenario and the standard deviation, skewness and kurtosis of the throughput time. Our analysis on the customer cases through these processes reveals that both expected and unexpected exceptions lead to a significantly longer throughput time ($p < 0.001$), thus proving hypothesis H1. Our analysis also shows that processes with expected exceptions and unexpected exceptions have significantly different throughput times ($p < 0.01$ in three disjoint Kruskal–Wallis H tests), thus proving hypothesis H2. The pairwise comparisons that were conducted separately for each process show that, when the exceptions are *unexpected*, the throughput times are significantly longer ($p < 0.002$ and $p < 0.012$, respectively for process P2 and process P5 scenario S2). Thus, paths involving unexpected exceptions have longer throughput times than those involving expected exceptions.

| Process | P2 – change control | P5 – incident management | |
|---|---|---|---|
| Scenario | S1-Handled | S1-Handled by Helpdesk | S2-Handled by others |
| # of instances analyzed | 3,982 | 1,844 | 4,742 |
| Avg. throughput time (days) | 51.6 | 0.39 | 16.2 |
| St. Deviation | 54.7 | 5.2 | 31.9 |
| Skewness | 3.09 | 37.45 | 7.83 |
| Kurtosis | 13.35 | 1537.83 | 77.17 |
| Expected Exceptions | ↑ | | ↑ |
| Unexpected Exceptions | ↑ | ↑ | ↑ |

LEGEND: ↑ : Increases throughput time (arrow size indicates degree of influence)
"empty" : Not applicable

**Table 5. The relation between expected/unexpected exceptions and the throughput time**

Table 6 presents the results from the statistical analyses performed over each process scenario and exception type (the results for each separate analysis is with $p < 0.01$), again including descriptive statistics of each scenario. Our analyses indicate that the throughput time in *all* process scenarios is statistically different when the paths involve different types of exceptions.

| | Process | P1 - appeals | | | | | P2 – change control | P3 – incident management | | P4 – loan application | | | P5 – incident management | |
|---|---|---|---|---|---|---|---|---|---|---|---|---|---|---|
| | Scenario | S1 Accepted | S2 Withdrawn | S3 Rejected | S4 Deferred | S6 Withdrawn & Rejected | S1 Handled | S1 Handled | S2 Cancelled | S1 Rejected | S2 Accepted | S3 Cancelled | S1 Handled Helpdesk | S2 Handled others |
| | # of instances analyzed | 775 | 173 | 158 | 90 | 39 | 3,982 | 39,547 | 628 | 8,034 | 2,244 | 2,807 | 1,844 | 4,742 |
| | Avg. throughput time (days) | 270.0 | 369.8 | 305.8 | 332.6 | 332.6 | 51.6 | 18.4 | 20.8 | 2.4 | 16.8 | 18.7 | 0.39 | 16.2 |
| | St. Deviation | 223.2 | 199.7 | 217.1 | 165.7 | 207.8 | 54.7 | 18.8 | 44.6 | 6.5 | 9.5 | 14.7 | 5.2 | 31.9 |
| | Skewness | .27 | -.37 | .08 | -.01 | -.01 | 3.09 | 6.61 | 4.48 | 3.85 | 1.86 | .58 | 37.45 | 7.83 |
| | Kurtosis | 1.31 | -1.12 | -1.52 | -.66 | -1.39 | 13.35 | 66.46 | 27.12 | 18.59 | 5.95 | .36 | 1537.83 | 77.17 |
| **EXCEPTION TYPE** | Early-Entry | | | | | | | | | | | | | ↔ (with Add & (with Repeat) |
| | Late-Entry | | | | | | | | | | | | ↑ | ↑ (with Add) |
| | Early-Exit | ↓ | | ↓ | | | | ↓ | | | | | | |
| | Late-Exit | | | ↓ (with Skip) | | | | ↓ (with Skip) | | ↑ (with Skip) | | | | |
| | Add | | ↓ (with Skip) | ↔ (with Skip) | ↑ | | ↑ | ↑ | ↑ | ↑ | | ↑ | ↑ (with Late-Entry) | ↑ |
| | Skip | | ↓ | ↓ (with Late-Exit) ↔ (with Add) | | ↓ | ↑ (with Add) | ↓ | | ↑ (with Late-Exit) | | ↓ | | |
| | Repeat | | | | | | ↑ | ↑ | ↑ | | | | | ↑ |
| | Step-Back | | | | | | ↑ (with Add) | ↑ | | ↑ | ↑ | ↑ | | |

**LEGEND:**
↑ : Longer throughput time
↓ : Shorter throughput time
↔ : No significant difference in throughput time
"empty" : Not applicable

**Table 6. Different types of exceptions and the throughput times** *(all with adjusted p < 0.01)*

The findings depicted in Table 6 can be aggregated over each exception type.

- In all five scenarios, where *Step-Back* appears, the throughput time is significantly longer than the normal flow.
- Similarly, in all scenarios that *Repeat* appears the throughput time is significantly longer.
- The analyses involving paths where *Skip* takes place, the throughput times differ significantly than the normal flow, but in different directions. In four scenarios in which it appears as the only exception type, the throughput times are shorter; whereas in two scenarios in which it appears paired with Add and Late-Exit exception types, the throughput times are longer. This can be attributed to the fact that the exception types that are paired with Skip (i.e. Add, Late-Exit) are expected to have process instances with longer throughput times (as discussed below). Hence, assuming that the exception type influences the throughput time, we may consider that the negative influence of the Skip in these scenarios was weaker when compared to the influence of Add and Late-Exit.
- *Add* appeared in majority of the scenarios that we analyzed (10 out of 13) - either separately or paired with other exception types. In eight of these scenarios, the throughput time is longer. Only in two scenarios, where the throughput time was shorter or remained the same, it appears with Skip, which is associated with shorter throughput times (as discussed above). Therefore, we may conclude that paths with Add exception types have significantly longer throughput time.
- *Late-Exit* occurred in three scenarios – always paired with Skip, the throughput times against the normal flow varied in different directions. Therefore, we do not have sufficient evidence to infer a valid conclusion on the relation between Late-Exit and the throughput time.
- In all scenarios where *Early-Exit* appears, the throughput time is significantly shorter than the normal flow.
- *Late-Entry* appears in two scenarios (in one, paired with Add). In both scenarios, the throughput time is significantly longer.
- *Early-Entry* appears (in sufficient numbers) only in the second scenario of Process 5, and paired with Add and Repeat. No significant relation is found between the presence of these pairs and the throughput time. However, as discussed above, our analysis on Add and Repeat shows that they are associated with increased throughput time. Therefore, we could argue that the insignificance of these pairs can be attributed to the significant negative influence of the Early-Entry exception, which balances the positive influence of the Add and Repeat exceptions and leads to an inconclusive result of the pairs' aggregated result. Nonetheless, this would still lead to a conclusion that is based on a single (qualitative) observation.

Based on the discussions above, we infer that the exception types Step-Back, Repeat, Add, and Late-Entry (i.e. exceptions that add activities to the normal flow) are associated with a significantly longer throughput time, while Skip and Early-Exit (i.e. exceptions that remove activities from the normal flow) are associated with a significantly shorter throughput time, thus proving hypothesis H3 and H4 to a large extent. For the Early-Entry and Late-Exit exception types, the result is inconclusive and further analysis is required with additional process data.

## 6    Conclusion

This paper has shown the relation between the occurrence of exceptions in a business process and the performance of that business process. It shows that the paths that host exceptions to the normal flow of a business process have higher throughput time in the overall. It also shows that unexpected exceptions (i.e. exceptions that are not taken into account in a pre-defined process model) are associated with a longer throughput time than expected exceptions (i.e. exceptions that are taken into account in a pre-defined process model). Finally, it investigates the relation between different types of exceptions and the throughput time, and shows that: paths that have exceptions that involve repeating one or more activities in a business process have longer throughput time. In addition, paths with exceptions that involve performing additional activities at the start or during a process have longer throughput time;

and paths with exceptions that involve skipping activities at the end or during a process have shorter throughput time.

These conclusions lead to practical implications on policies that can be followed with respect to exceptions. First, since expected exceptions are associated with shorter throughput time than unexpected exceptions, it is advisable to investigate in advance which exceptions may occur during a process. These exceptions should then also be considered when developing operational support, such as forms, work instructions and information systems. However, there is a trade-off to be made here, because processes typically have too many paths to all be considered up front. Nonetheless, some of the processes that we studied contained unexpected exceptions within the 10 most frequently occurring paths, so improvement seems possible. Second, not all exceptions can be related to a longer throughput time. Anecdotal evidence suggests that employees sometimes take shortcuts in order to improve on process performance. For example, another study (Martens, 2009) showed that employees sometimes skipped a mandatory 'resource reservation' step and instead agreed with their colleagues that they could use the resource in order to meet a deadline and keep customers happy. However, we also saw situations in which employees deliberately skipped steps with a detrimental effect. Therefore, employees should be enabled and even encouraged to make exceptions to the normal flow if necessary, but should also be monitored to detect mistakes and prevent those mistakes in the future.

It is important to note that in this study correlation certainly does not equal causation. For example, in one process we found that the additional activity 'send reminder' was related to a longer throughput time (Jager, 2015). However, it was of course not the 'send reminder' activity that caused the longer throughput time. It was rather the fact that the customer was unresponsive that caused both the longer throughput time and the reminder to be sent. This has the important practical implication that no general rules can be created for how to handle exceptions, but that each exception has to be studied by itself to determine how to mitigate its effects.

The research is performed by studying the customer cases of five processes from five different companies. First, it describes a statistical analysis of the relations between exceptions and performance per process, as there are enough customer cases per process to conduct a statistical analysis. Subsequently, it qualitatively aggregates over the results of the individual processes to derive general conclusions. The results of this analysis confirm the expectations regarding the relation between control-flow exceptions and throughput time. To the best of our knowledge, this work is the first to empirically test these expectations and provide quantitative evidence for them.

From a methodological perspective the two most important limitations are the following. First, this paper defines an exception as any path through a process that is not the normal flow of a scenario. A scenario is a particular outcome of a process, such as 'accepted', 'rejected', or 'sent on to other department'. The normal flows through these scenarios can safely be defined as non-exceptional. Also, in all processes that are studied, there is a large number of paths that can safely be defined as exceptional, because they are taken by less than 1% of the customer cases. However, there also is a grey area of customer cases that could both be classified as 'normal' and as 'exceptional'. We chose a rigorous definition of 'exception', but asking domain experts to identify paths as either 'normal' or 'exceptional' would yield a set of exceptions that is closer to a practical interpretation of exception. In future work, we aim to ask practitioners to identify their exceptions and investigate whether that leads to significantly different results. Second, due to limitations of the available data, this paper only studies the performance of processes in terms of throughput time with the presumption that favors a shorter time. In future work, we also aim to study the effect of exceptions on other performance indicators to provide a better representation of operational performance. Of particular interest are customer satisfaction and cost. We conducted our analyses over 70.000 cases originating from five industry processes. However, future work should consider analyzing instances of additional processes to strengthen the generalizability of the results.

# References

Adams, M. J., ter Hofstede, A. H., Edmond, D. and van der Aalst, W.M.P. (2006). "Worklets: A Service-Oriented Implementation of Dynamic Flexibility in Workflows." In: *Proceedings of CoopIS*. Berlin-Heidelberg: Springer, p. 291-308.

Curbera, F., Khalaf, R., Leymann, F. and & Weerawarana, S. (2003). "Exception handling in the BPEL4WS language." In: *Proceedings of BPM*. Berlin-Heidelberg: Springer, p. 276-290.

Davenport, T. (1993). *Process Innovation: Reengineering work through information technology*. Boston: Harvard Business School Press.

Eder, J. and Liebhart, W. (1998). "Contributions to exception handling in workflow management." In: *Proceedings of the EDBT Workshop on Workflow Management*. p. 3-10.

Gottschalk, F., Wagemakers, T.A., Jansen-Vullers, M.H., van der Aalst, W.M.P. and La Rosa, M. (2009). "Configurable process models: Experiences from a municipality case study." In: *Proceedings of CAiSE*. Berlin-Heidelberg: Springer, p. 486-500.

Grigori, D., Casati, F., Castellanos, M., Dayal, U., Sayal, M. and Shan, M.-C. (2004). "Business Process Intelligence." *Computers in Industry* 53(3), 321-343.

Grigori, D., Casati, F., Dayal, U. and Shan, M.-C. (2001). "Improving Business Process Quality through Exception Understanding, Prediction, and Prevention." In: *Proceedings of VLDB*. San-Francisco: Morgan Kaufmann, 159-168.

Field, A. (2013). *Discovering Statistics using IBM SPSS Statistics*. Washington: SAGE Publications Ltd.

Hagen, C. and Alonso, G. (2000). "Exception handling in workflow management systems." *IEEE Transactions on Software Engineering* 26(10), 943-958.

Hallerbach, A., Bauer, T. and Reichert, M. (2010). "Capturing variability in business process models: the Provop approach." *Journal of Software Maintenance and Evolution: Research and Practice* 22(6-7), 519-546.

Hwang, S.Y., Ho, S.F. and Tang, J. (1999). "Mining exception instances to facilitate workflow exception handling." In: *Proceedings of DASFAA*. Los Alamitos: IEEE Computer Society, p. 45-52.

Martens, A. C. N. (2009). *Relevance of conformance analysis information*. Master's Thesis. Eindhoven: Eindhoven Univerity of Technology.

Jager, L. de (2015). *Consequences of Exceptions in Business Processes*. Master's Thesis. Eindhoven: Eindhoven University of Technology.

Meerkamm, S. (2012). "Staged configuration of multi-perspectives variants based on a generic data model." In: *Proceedings of the BPM Workshops*. Berlin-Heidelberg: Springer, p. 326-337.

Object Management Group (2011). *Business Process Model and Notation (BPMN) version 2.0*. Technical report formal/2011-01-03. Object Management Group.

Reichert, M. and Weber, B. (2012a). "AristaFlow BPM Suite." In: *Enabling Flexibility in Process-Aware Information Systems*. Berlin-Heidelberg: Springer, p. 441-464.

Reichert, M. and Weber, B. (2012b). *Enabling flexibility in process-aware information systems: challenges, methods, technologies*. Berlin-Heidelberg: Springer.

Reichert, M. and Dadam, P. (1998). "ADEPTflex—Supporting dynamic changes of workflows without losing control." *Journal of Intelligent Information Systems* 10(2), 93-129.

Rinderle, S. and Reichert, M. (2006). "Data–driven process control and exception handling in process management systems." In: *Proceedings of CAiSE*. Berlin-Heidelberg: Springer, p. 273-287.

Rozinat, A. and van der Aalst, W.M.P. (2008). "Conformance checking of processes based on monitoring real behavior." *Information Systems* 33(1), 64-95.

Russell, N., van der Aalst, W.M.P. and ter Hofstede, A. (2006). "Workflow Exception Patterns." In: *Proceedings of CAiSE*. Berlin-Heidelberg: Springer, p. 288-302.

Schonenberg, H., Mans, R., Russell, N., Mulyar, N. and van der Aalst, W.M.P. (2008). "Process flexibility: A survey of contemporary approaches." In: *Advances in Enterprise Engineering I*. Berlin-Heidelberg: Springer, p. 16-30.

Steeman, W. (2013). *Log of Volvo IT incident management system.* DOI: 10.4121/uuid:500573e6-accc-4b0c-9576-aa5468b10cee (visited on 11/13/2014).

Strong, D. M. (1997). "IT process designs for improving information quality and reducing exception handling: a simulation experiment." *Information & Management* 31(5), 251-263.

Tibshirani, R., Walther, G. and Hastie, T. (2001). "Estimating the number of clusters in a data set via the gap statistic." *Journal of the Royal Statistical Society: Series B* 63, 411–423

van der Aalst, W.M.P. (2009). "Process-aware information systems: Lessons to be learned from process mining." In: *Transactions on petri nets and other models of concurrency II*. Berlin-Heidelberg: Springer, p. 1-26.

van Dongen, B.F. (2012). *Event log of a loan application process*. DOI:10.4121/uuid:3926db30-f712-4394-aebc-75976070e91f (visited on 11/13/2014).

Weber, B., Reichert, M., Mendling, J. and Reijers, H. A. (2011). "Refactoring large process model repositories." *Computers in Industry* 62(5), 467-486.

Weber, B., Reichert, M. and Rinderle-Ma, S. (2008). "Change patterns and change support features–enhancing flexibility in process-aware information systems." *Data & knowledge engineering* 66(3), 438-466.

Witte, R. S. and Witte, J. S. (2014). *Statistics*. 10th Edition. Hoboken: John Wiley & Sons.

Zeng, L., Lei, H., Jeng, J.J., Chung, J.Y. and Benatallah, B. (2005). "Policy-driven exception-management for composite web services. In E-Commerce Technology." In: *Proceeding of CEC*. Los Alamitos: IEEE Computer Society, p. 355-363.

zur Muehlen, M. and Recker, J. (2008). "How much language is enough? Theoretical and practical use of the business process modeling notation." In: *Proceedings of CAiSE*. Berlin-Heidelberg: Springer, p. 465-479.